\documentclass{IEEEtran}
\usepackage{cite}
\usepackage{amsmath,amssymb,amsfonts}
\usepackage{xcolor}
\usepackage{algorithmic}
\usepackage{graphicx}

\usepackage[utf8]{inputenc}
\usepackage[english]{babel}
\usepackage{fancyhdr}
\usepackage{hyperref}
\usepackage{textcomp}
\def\BibTeX{{\rm B\kern-.05em{\sc i\kern-.025em b}\kern-.08em
    T\kern-.1667em\lower.7ex\hbox{E}\kern-.125emX}}
\begin{document}
\title{{\fontsize{24}{26}\selectfont{Communication\rule{29.9pc}{0.5pt}}}\break\fontsize{16}{18}\selectfont
Body-UAV Near-Ground LoRa Links through a Mediterranean Forest}
\author{Giulio Maria Bianco, \IEEEmembership{Member, IEEE} and Gaetano Marrocco \IEEEmembership{Senior Member, IEEE}
\thanks{Submitted for review on XX Month 20XX.}
\thanks{G. M. Bianco is with the Department of Civil Engineering and Computer Science Engineering of the University of Rome Tor Vergata, Rome, Italy (e-mail: giulio.maria.bianco@uniroma2.it).}
\thanks{G. Marrocco is with the Department of Civil Engineering and Computer Science Engineering of the University of Rome Tor Vergata, Rome, Italy (e-mail: gaetano.marrocco@uniroma2.it).}}

\maketitle

\begin{abstract}
LoRa low-power wide-area network protocol has recently gained attention for deploying ad-hoc search and rescue (SaR) systems. They could be empowered by exploiting body-UAV links that enable communications between a body-worn radio and a UAV-mounted one. However, to employ UAVs effectively, knowledge of the signal's propagation in the environment is required. Otherwise, communications and localization could be hindered. The radio range, the packet delivery ratio (PDR), and the large- and small-scale fading of body-UAV LoRa links at $\mathbf{868}$~MHz when the radio wearer is in a Mediterranean forest are here characterized for the first time with a near-ground UAV having a maximum flying height of $\mathbf{30}$~m. A log-distance model accounting for the body shadowing and the wearer's movements is derived. Over the full LoRa radio range of about $\mathbf{600}$~m, the new model predicts the path loss (PL) better than the state-of-the-art ones, with a reduction of the median error even by $\sim\mathbf{10}$~dB. The observed small-scale fading is severe and follows a Nakagami-m distribution. Extensions of the model for similar scenarios can be drawn through appropriate corrective factors.
\end{abstract}


\section{Introduction}\label{sec:Introduction}
LoRa is a convenient low-power wide-area network protocol that can enable kilometric communications by low transmission power~\cite{Cheriyan22,Ameloot21,Bianco21LoRa}. It is hence extremely promising for fostering search and rescue (SaR) missions~\cite{Bouras21} that usually take place in harsh environments like mountains~\cite{Bianco20Performance}, disaster areas~\cite{Manuel22}, and forests~\cite{Calabro21}. Forested areas, in particular, are challenging scenarios for LoRa devices due to the detrimental effects of vegetation on path loss (PL)~\cite{Seybold05,Ahmad18}. Indeed, the radio range is lowered down to a few hundred meters independently from the selected transmission parameters~\cite{Sharma22, Hakim22, Ferreira20}. Indirect propagation can even be forbidden~\cite{Callebaut20}, and the PL heavily depends on the density, type, and age of the vegetation~\cite{Guerriero20}. \par
To the authors' best knowledge, even though terrestrial LoRa links in forests were already investigated, air-ground (AG) links in such environments have never been considered. Instead, unmanned aerial vehicles (UAVs) and systems (UASs) are game-changing technologies for SaR missions. They could dramatically increase the survival chances of the targets to be rescued by avoiding terrestrial obstacles and speeding up operations through near-ground flights having altitudes of tens of meters \cite{Xing22, Grau21}. Thus, body-UAV links, viz., electromagnetic links between a body-worn device and a UAV~\cite{Kachroo19}, can maximize the benefits of using wearable LoRa devices~\cite{Bianco22Numerical}. However, available ground-UAV propagation models are not suitable for describing the signal attenuation in forests since they are tuned for urban environments mainly and do not account for the presence and movements of the radio wearer.\par
\begin{figure}[tp]
  \centering
  \includegraphics[width=8cm]{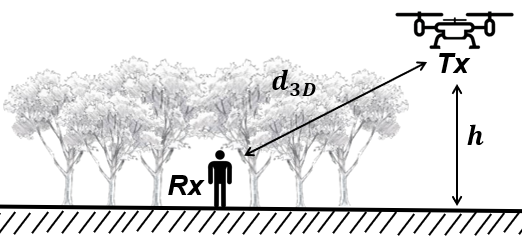}
  \caption{Sketch and parameters of the propagation scenario for a body-UAV LoRa communication through a Mediterranean forest.}
  \label{fig:forestSketch}
\end{figure}

By expanding the preliminary results in~\cite{Bianco22Indirect}, this paper presents for the first time the characterization of low-altitude body-UAV LoRa non-line-of-sight~(NLoS) links through a Mediterranean forest~(Fig.~\ref{fig:forestSketch}). In the considered scenario, the user wearing a LoRa receiver is moving inside the forest, while a SaR quadcopter, equipped with a LoRa transmitter, is hovering just outside the forest. Near-ground altitudes ($h\leq30$~m) are considered to reproduce fast, on-site deployment of a low-cost drone that broadcasts emergency signals, can be carried by terrestrial SaR teams, and could swiftly become unstable, displace or discharge due to turbulent winds if its altitude is not low enough \cite{Wang19}. As body-UAV links combine AG and off-body links, the derived empirical model includes the equivalent gain of the on-body antenna and the polarization losses, and it is compared with existing AG models.\par
\section{Methods and Instrumentation}
\subsection{Body-UAV Propagation Modeling}\label{subsec:modeling}
The PL of channels involving a terrestrial radio and a low-flying UAV is here modelled by a log-distance representation which depends on the flying height of the UAV \cite{Cui19, 3GPPTR38901}. This approach avoids the use of a reference terrestrial PL~\cite{AlHourani18}. Furthermore, off-body links need to account for the variable Tx-Rx angular arrangements and the movements of the radio's wearer. The gain of the body-worn antenna and the polarization losses are hereafter taken into account using a numerical-statistical approach \cite{Bianco22Numerical}, as detailed in \S\ref{subsec:equivalentGPLF}, whereas the movements can cause unpredictable body shadowing and polarization losses and, thus, additional fluctuations of the signal's strength.\par
With reference to Fig.~\ref{fig:forestSketch}, for a given frequency and indirect propagation, the AG link is parametrized \cite{Cui19} as
\begin{equation} \label{eq:PLUAV}
PL_m\left(d_{3D},h\right)=PL_m\left(d_0\right)+10\gamma\cdot\log_{10}\left(d_ {3D}\right) - \eta h,
\end{equation}
where $PL_m$ is the mean PL, $d_{3D}$ is the Tx-Rx tridimensional distance between man and UAV, $h$ is the UAV flying height, $d_0$ is a reference distance in the far-field of the transmitter ($d_0=1$~m in the following), $\gamma$ is the path loss exponent, and $\eta$ is the altitude impact factor. The parameters fitting the experimental data are hence $\left\{PL\left(d_0\right), \gamma, \eta\right\}$.\par
Small- and large-scale fluctuations are accounted for by the \textit{instantaneous} path loss~($PL_i$). The small-scale fading $f_{sm}$ includes the fluctuations within the order of a wavelength~$\lambda$~\cite{Seybold05} and can be filtered out by means of a $\pm\lambda/2$~moving window average~\cite{Cui19}. $PL_m$ is derived by application of least-square fitting to the PL measurements through~(\ref{eq:PLUAV}) after the removal of the small-scale fading. Finally, a zero-mean gaussian shadow fading with standard deviation $\sigma_{SF}$~\cite{Cui19} models the remaining large-scale differences between the instantaneous and the mean PL. The $PL_i$ is hence written as
\begin{equation} \label{eq:InstantaneousPL}
PL_i\left(d_{3D},h\right)=PL_m\left(d_{3D},h\right) + f_{sm} + \mathcal{N}\left(0,\sigma_{SF}\right).
\end{equation}
$PL_i\left(d_{3D},h\right)$ is derived from the received signal strength indicator (RSSI) and signal-to-noise ratio (SNR) recorded by the receiver LoRa board \cite{Bianco22Numerical,Bianco21LoRa} as
\begin{multline}  \label{eq:ExperimentalPL}
PL_i\left(d_{3D},h\right)=P_{Tx} + G_{Tx} + G_{Rx} + \chi + \\ + 10\cdot\log_{10}\left(1+\frac{1}{SNR}\right)-RSSI
\end{multline}
being $P_{Tx}$ the transmission power, $\chi$ the polarization loss factor (PLF) between the Tx and Rx antennas, and $G_{Tx}$ and $G_{Rx}$ the corresponding radiation power gains. The PL calculated in (\ref{eq:PLUAV}) and (\ref{eq:InstantaneousPL}) and all the terms in (\ref{eq:ExperimentalPL}) are exposed in the dB scale except for the SNR, which is in the linear scale. \par
Since the resulting model fits experimental data, it will be accurate under the same environmental conditions as during the data collection. Nevertheless, different conditions that can affect the signal's strength and increase the difference with the expected PL can be accounted for by an appropriate corrective factor $\xi$ of the $PL_m$ computed according to (\ref{eq:PLUAV}). The refined estimation of the median path loss~($PL_m'$) can be hence expressed as
\begin{equation} \label{eq:PLUAVCorretto}
PL_m'\left(d_{3D},h\right)=PL_m\left(d_{3D},h\right)+\xi.
\end{equation}
The corrective factor can be parametrized regarding the difference of RSSI~($\Delta RSSI$) and SNR~($\Delta SNR$) between the values in the new conditions and the reference ones
\begin{multline}  \label{eq:correctiveFactor}
\xi\left(\Delta RSSI,SNR,\Delta SNR\right)=-\Delta RSSI+\\ 
+10\cdot\log_{10}\left[1+10^{-\left(SNR+\Delta SNR\right)/10}\right]
\end{multline}
where all the terms in (\ref{eq:PLUAVCorretto}) and (\ref{eq:correctiveFactor}) are in the dB scale. As reported by the literature on LoRa, $\left\{\Delta RSSI, \Delta SNR\right\}$ could depend on the type of vegetation~(e.g., Malaysian palm~\cite{Anzum22}, Japanese mountainous forest~\cite{Myagmardulam21}, eastern China mixed forest~\cite{Wu20AModel}), season~\cite{Sudirjo19,Ameloot21Variable,Elijah21}, weather~\cite{Elijah21,Parri21,Gaelens17} and the relative speed between the antennas yielding to significant Doppler effect~\cite{Ameloot21Characterization,Petajajarvi17}.
\begin{figure}[tbp]
  \centering
\qquad \includegraphics[height=34mm]{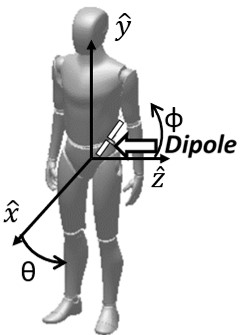}\qquad\qquad\includegraphics[height=34mm]{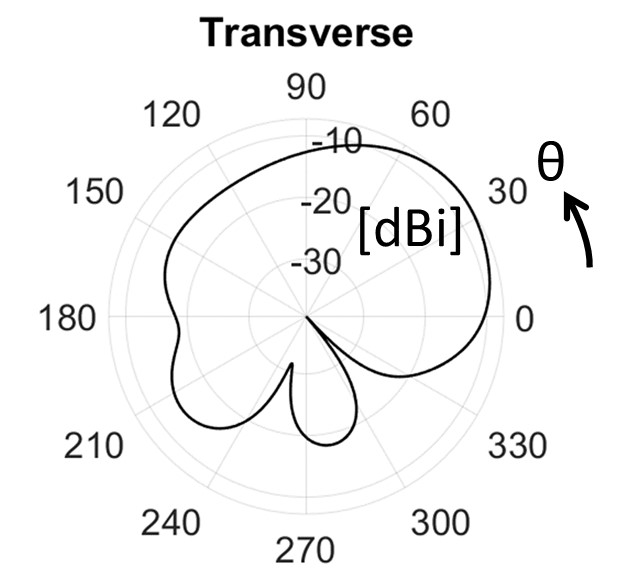}\\
(a) \qquad \qquad \qquad \qquad \qquad \qquad (b)\\
  \includegraphics[height=34mm]{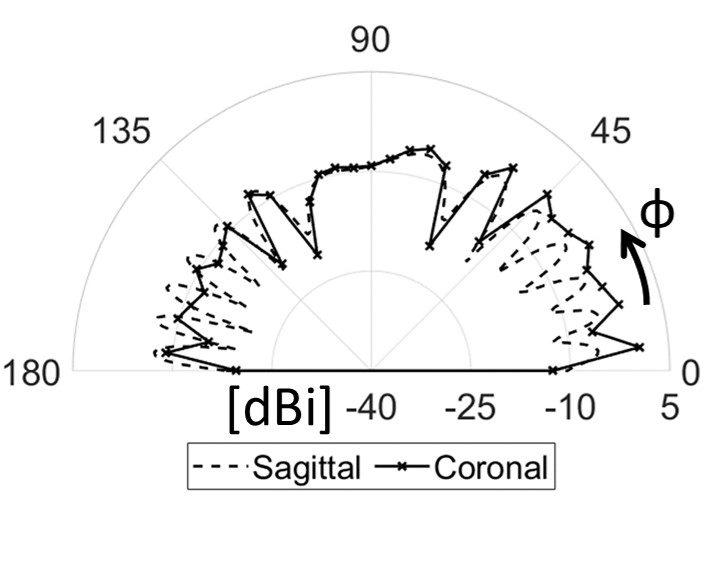} \includegraphics[height=34mm]{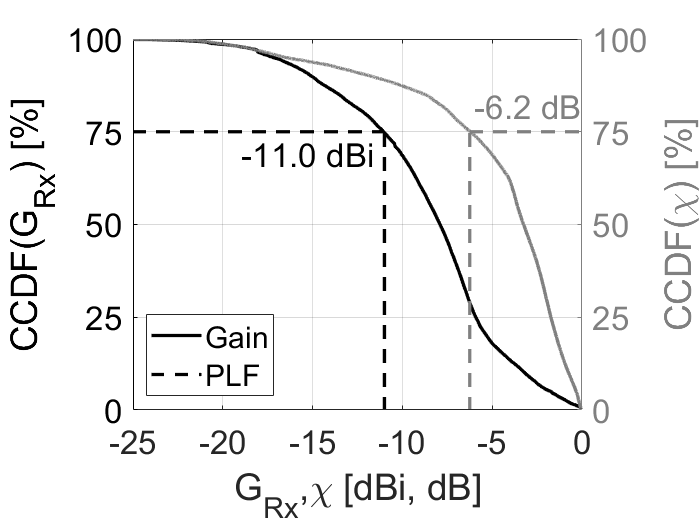}\\
(c) \qquad \qquad \qquad \qquad \qquad \qquad (d)\\
  \caption{Numerical simulations of the on-body receiver. (a) Phantom of the body-worn antenna with the coordinate reference system. Polar plots of the receiver's gain according to the anatomical planes: (b) transverse and (c) sagittal and coronal planes. (d) CCDF of $G_{Rx}$ and the PLF.}
  \label{fig:equivalentGainPLF}
\end{figure}
\subsection{Equivalent Gains and Polarization Losses}\label{subsec:equivalentGPLF}
%
The computation of (\ref{eq:ExperimentalPL}) requires estimating the parameters $\{G_{Tx},G_{Rx},\chi\}$. Under the assumption that the radiation patterns of the two antennas are not very directive and that the range of the relative movements between them is limited, the gains of the transmitter and the receiver are considered invariant with respect to their mutual orientation and distance. In particular, the transmitter's gain is fixed to $G_{Tx}=0$~dBi since the position of the antenna onboard the UAV can be properly optimized to provide a reasonable and uniform gain towards the ground. The receiver's effective gain and the PLF
\begin{equation} \label{eq:PLFEvaluation}
\chi=\left|\hat{\rho}_{Tx}\cdot\hat{\rho}_{Rx}^*\right|^2
\end{equation}
with `$^*$' the conjugate operator, `$\cdot$' the inner product, and $\hat{\rho}_{Tx/Rx}$ the unitary polarization vector of the Tx/Rx, are instead estimated through a statistical approach. The method is conservative and accounts for the body shadowing by numerical simulations\footnote{Numerical simulations performed through CST Microwave Studio $2018$.} involving a homogeneous anatomical phantom of the human body\footnote{The model is available at https://grabcad.com/library/corpo-humano-masculino-adulto-by-jari-ikonen-rev-01-1.}. The phantom is made of muscle-like material (conductivity~$\sigma=1.56$~S/m\textsuperscript{-1}; relative permittivity~$\epsilon_r=51.6$; density~$1055$ kg/m\textsuperscript{$3$} \cite{Scanlon00, Ward05}) and stands on an infinite, perfect electric conductor (PEC) ground. The simulated LoRa dipole (a $1$-mm-wide and $35$-$\mu$m-thick copper trace long $155$~mm; Fig.~\ref{fig:equivalentGainPLF}(a)) is placed at $5$~mm from the skin, as it was in a pocket during the experiments. Both the transmitting and receiving dipoles form a $30^{\textnormal{o}}$-angle with the planes parallel to the ground so that \textit{i})~nulls in the radiation pattern of the dipoles are not expected along the ray path, \textit{ii})~the dipoles' axes stay parallel for maximum polarization coupling, and \textit{iii})~movements of the UAV's rotors are not hindered.\par
\begin{table}[t]
  \centering
  \caption{Employed LoRa's transmission parameters.}
\begin{tabular}{l|l||l|l}\label{tab:LoRaParameters}
  \bf{Transmission} & & \bf{Transmission} & \\
\bf{parameter} & \bf{Value} & \bf{parameter} & \bf{Value} \\
 \hline\hline
 Transmission power & $14$~dBm & Bandwidth & $125$~kHz \\
\hline
 Carrier frequency &$868$~MHz & Coding rate &$4/5$\\
\hline
 Spreading factor & $7$ & Message rate &$4^{-1}$ msg/sec\\
\end{tabular}
\end{table}
\begin{figure}[tp]
  \centering
  \includegraphics[height=24mm]{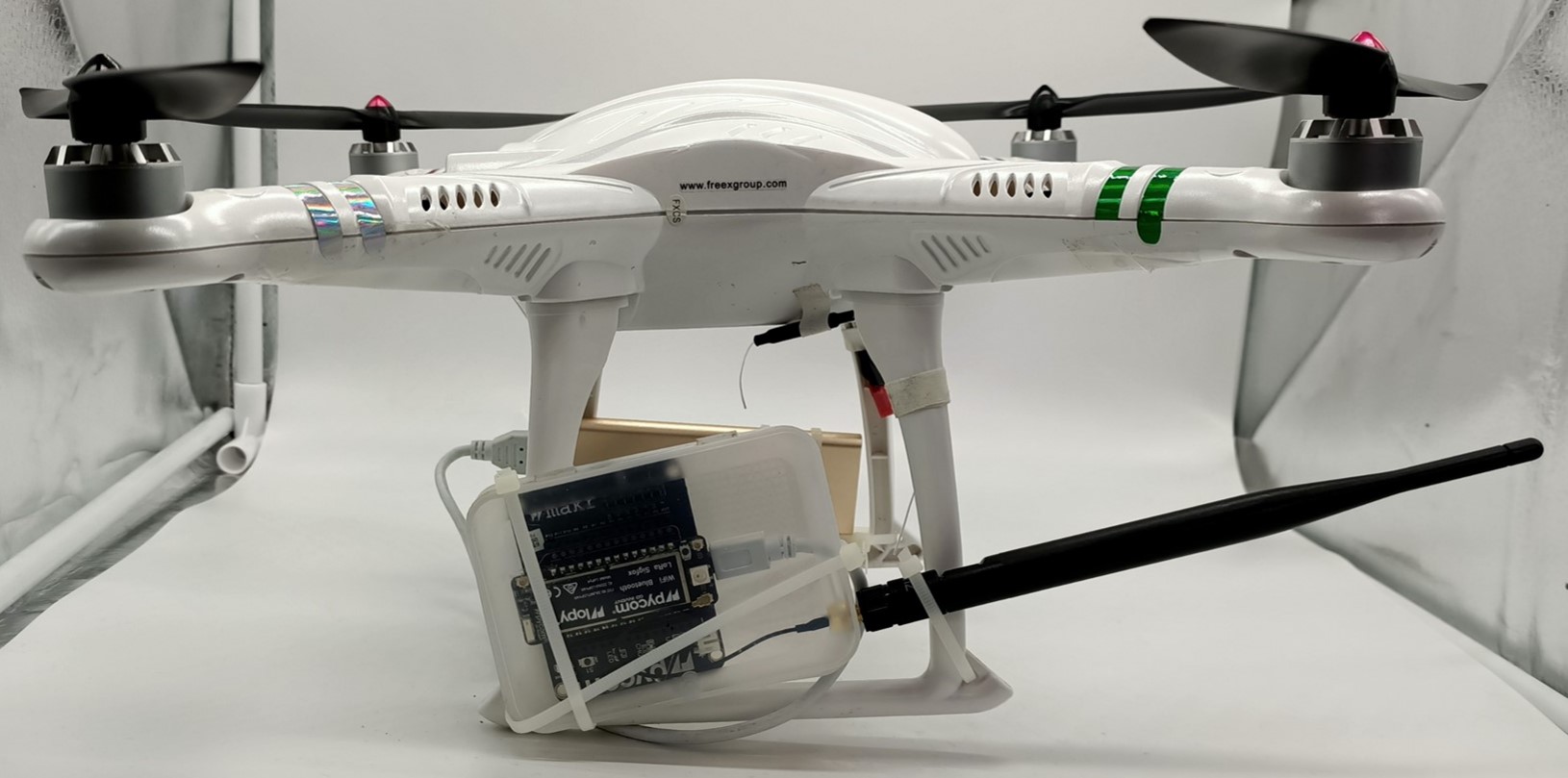}  \includegraphics[height=24mm]{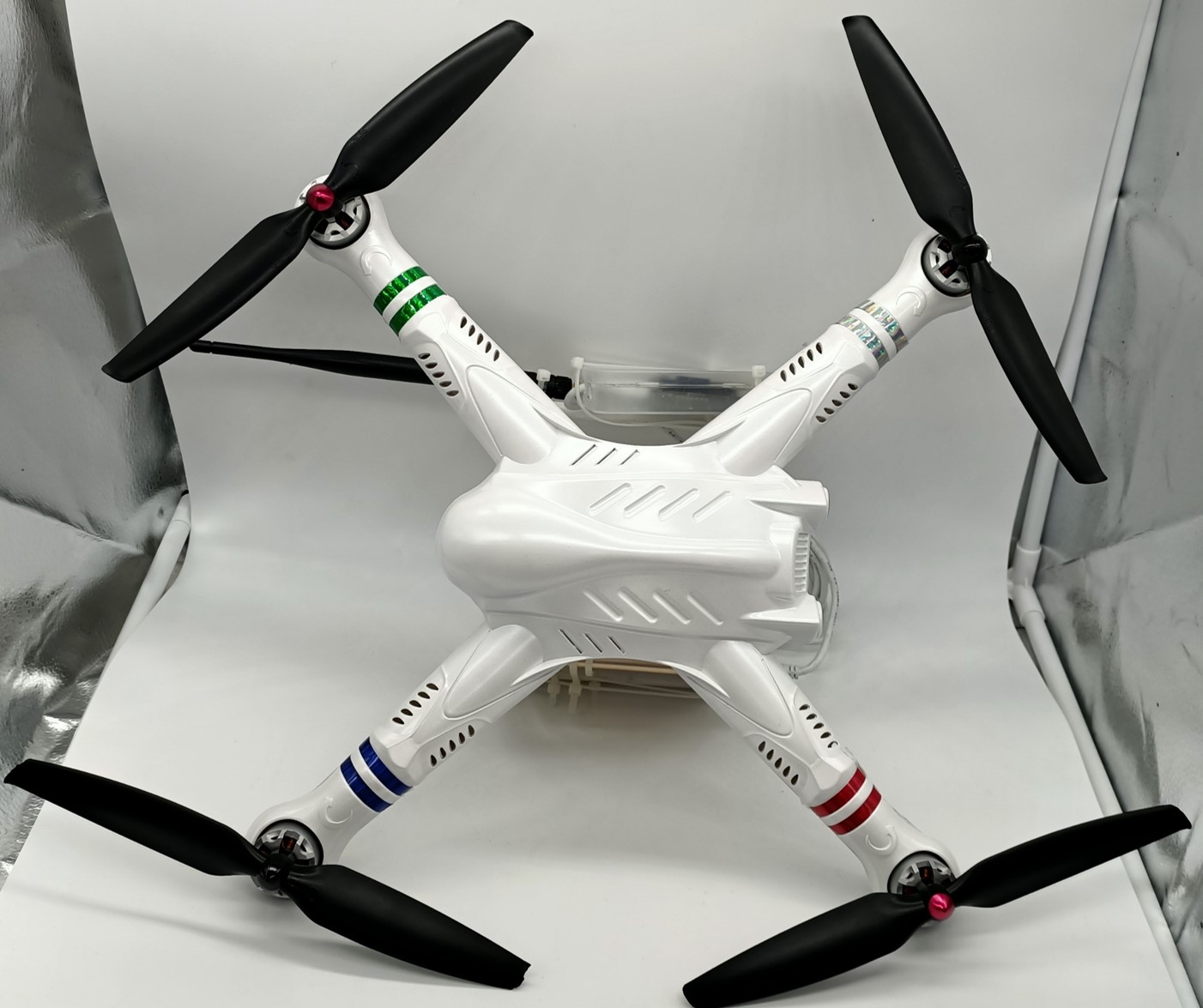}
  \begin{tabular}{p{5cm}p{3cm}}
  \qquad \qquad \qquad (a) & \qquad (b)
  \end{tabular}
  \includegraphics[width=80mm]{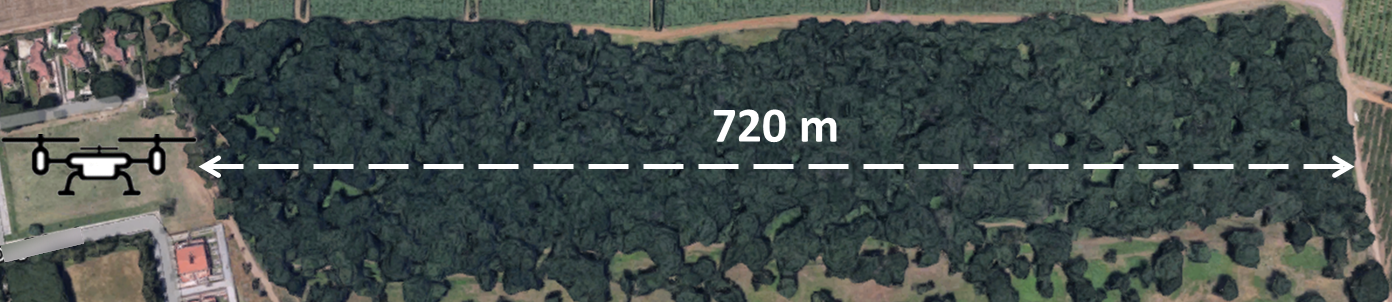}\\
(c)\\
    \caption{UAV equipped with the LoRa transmitter: (a) side and (b) top view. (c) Satellite view of the forested area. The position of the hovering UAV and the path walked by the volunteer are highlighted.}
  \label{fig:hardwareAndTest}
\end{figure}
\begin{table}[t]
  \centering
  \caption{Maximum observed communication range and PDRs for the three flying heights of the UAV.}
\begin{tabular}{l||l|l|l}\label{tab:WoodRangePDR}
  & $\mathbf{h=3}$~\bf{m} & $\mathbf{h=10}$~\bf{m} & $\mathbf{h=30}$~\bf{m} \\
 \hline\hline
 \bf{Radio range} & $597$~m & $676$~m & $589$~m \\
\hline
 \bf{PDR} &$93.7\%$ &$91.9\%$ &$91.0\%$\\
\end{tabular}
\end{table}
\begin{table*}
  \centering
  \caption{Path loss model parameters for body-UAV links.}  \label{tab:WoodModel}
\begin{tabular}{l||l|l|l|l|l} & \bf{Mediterranean} & \bf{3GPP UMa} & \bf{3GPP UMa} & \bf{3GPP UMi} & \bf{Cui's NLoS}\\
 \bf{Model} & \bf{forest (this work)} & \bf{TR 38.901 \cite{3GPPTR38901}} & \bf{TR 38.901 optional \cite{3GPPTR38901}} & \bf{TR 38.901 \cite{3GPPTR38901}} & \bf{($\mathbf{1}$ GHz) \cite{Cui19}}\\
 \hline\hline
 \bf{PL exponent} $\gamma$ & $4.19$ & $3.91$ & $3.00$ & $3.53$ & $2.00$ \\  
 \hline
\bf{PL intercept} $\mathbf{PL(1}$ \, \bf{m}$\mathbf{)}$ &  $10.69$ dB & $13.21$ dB & $31.17$ dB & $21.54$ dB & $61.18$ dB \\
 \hline
\bf{Altitude impact factor} $\eta$ &$0.12$ $\textnormal{dB}\left(\textnormal{m}\right)^{-1}$ & $0.60$ $\textnormal{dB}\left(\textnormal{m}\right)^{-1}$ & Absent & $0.30$ $\textnormal{dB}\left(\textnormal{m}\right)^{-1}$ & $1.19$ $\textnormal{dB}\left(\textnormal{m}\right)^{-1}$\\
 \hline
 \bf{Shadow fading} $\sigma_{SF}$ & $8.05$ dB & $6.00$ dB & $7.80$ dB & $7.82$ dB & $3.60$ dB \\
 \hline\hline
 \bf{Median difference} &&&& \\
 \bf{with measurements} & $\mathbf{7.34}$ \textbf{dB}&$14.96$ dB&$11.44$ dB&$13.37$ dB&$17.00$ dB  \\
\end{tabular}
\end{table*}
$G_{Rx}$ and $\chi$ are computed over the half-space free from the PEC ground (i.e., $0^{\textnormal{o}}\leq\phi\leq180^{\textnormal{o}}$) with a $1^{\textnormal{o}}$ resolution. Fig.~\ref{fig:equivalentGainPLF}(b,c) show the polar plots of the gain over the three principal anatomical planes. Then, the effective values to be used in (\ref{eq:ExperimentalPL}) are derived from complementary cumulative distribution functions (CCDFs) as the minimum value that is guaranteed in $75\%$ of the Tx-Rx angular arrangements. The receiver's versor is obtained from simulations, whilst constant $\hat{\rho}_{Tx}=-\cos\left(30^{\textnormal{o}}\right)\hat{x}+\sin\left(30^{\textnormal{o}}\right)\hat{y}$ is considered for the transmitter. Overall, the values $G_{Rx}\left(CCDF=75\%\right)=-11.0$~dBi and $\chi\left(CCDF=75\%\right)=-6.2$~dB~(Fig.~\ref{fig:equivalentGainPLF}(d)) are finally inserted in (\ref{eq:ExperimentalPL}) to derive $PL_i$. The resulting PL model will account for the residual variability that is not captured by this approximation.
\subsection{Link Characterization}\label{subsec:linkchara}
The link is characterized based on \textit{i}) the maximum communication range, \textit{ii}) the packet delivery ratio (PDR; the ratio between received and sent packets), \textit{iii}) the large-scale fading and \textit{iv}) the small-scale fluctuations. The large-scale fading comprises the mean PL and the large-scale fluctuations.\par
As mentioned above, this work considers the rapid changes of the signal strength over a short distance on the order of a wavelength~\cite{Seybold05,Grami15} as small-scale fading. This is a crucial phenomenon for predicting outage probability since such fading can be as high as $40$~dB~\cite{Grami15} and can follow several different statistical distributions. According to \cite{Cui19}, the small-scale fluctuations are analyzed with a spatial moving average and by fitting its empirical statistical distribution using a set of trial fits, viz., log-logistic, Nakagami-m, Weibull, Rayleigh, and Rician. The fade depth is the difference (in decibel scale) of small-scale signal fading between the $50$\% and $99$\% levels, which are computed as in \cite{He13}, and denotes the possible expected fluctuations.\par
The statistical validity of the estimated PL (EPL) model is verified by calculating the relative standard error of the mean (RSE) for each $\{d_{3D},h\}$ pair as~\cite{AlHourani18}
\begin{equation} \label{eq:RSE}
\scalebox{0.85}{
$RSE_{PL_m}\left(d_{3D},h\right)=\sigma_{SF}\left[\sqrt{M\left(d_{3D},h\right)}PL_m\left(d_{3D},h\right)\right]^{-1}$}
\end{equation}
being $M\left(d_{3D},h\right)$ the number of $PL_i\left(d_{3D},h\right)$ measurements averaged, after subtracting the small-scale fading, to obtain a measurement of the large-scale PL for each pair. \par
Finally, to assess the improvement in accuracy of the PL prediction, the derived EPL is compared with known AG models suitable for LoRa indirect propagation. The considered models are \textit{i}) 3GPP urban macrocell (UMa) NLoS~\cite{3GPPTR38901}, \textit{ii}) 3GPP optional UMa NLoS model \cite{3GPPTR38901}, \textit{iii}) 3GPP optional urban microcell (UMi) \cite{3GPPTR38901}, and \textit{iv}) Cui's NLoS model at $1$~GHz \cite{Cui19}. These models that describe large-scale fading are optimized for urban scenarios and assume a fixed terrestrial radio. The accuracy of each model is evaluated by considering the difference between the EPL and $PL_i$.
\subsection{Measurement Set-Up}
The experimental LoRa system comprises two LoPy-4 boards (by Pycom; datasheet available online~\cite{LoPy417}) connected to commercial half-wavelength dipoles. The LoRa transmission parameters (Table~\ref{tab:LoRaParameters}) maximize the RSS seen by the UAV and the message rate in compliance with the European duty-cycle limitations \cite{Bianco21LoRa}. The UAV is a low-cost quadcopter for fast deployment (FreeX model by Only Flying Machines). The transmitting board is fixed to the quadcopter~(Fig.~\ref{fig:hardwareAndTest}(a,b)), while the body-worn receiver is tied in the pocket of the jacket of a volunteer. The dipoles form the same $30^{\textnormal{o}}$-angle with the ground as in the simulations.\par
The forest for the experiment is in Colle Romito (Ardea, Lazio, Italy; maximum extension $\sim720$~m; GPS coordinates $41^{\textnormal{o}}33'01.9"$N $12^{\textnormal{o}}35'08.4"$E) and is mostly composed of poplars (scientific name "Populus tremula"). The trees are young adults, not pruned, and approximately $10$~m tall. All measurements were completed on sunny days (relative humidity~$\sim30$\%, temperature~$\sim25$~Celsius). The Doppler effect was neglected as the maximum relative speed between the antennas was about $1$~m/s. Overall, about $1300$ data were collected during the measurements.\par
The transmitting UAV hovered at the limit of the forest, at a fixed flying height, without moving. In this way, there is some flexibility in enforcing the hovering height when transmitting emergency signals to warn and group people in the area~\cite{Bianco21LoRa}, e.g., in the case of wildfires. The GPS-tracked volunteer was slowly walking along a roughly straight path~(Fig.~\ref{fig:hardwareAndTest}(c)) inside the forest. Trees, trunks, and other obstacles forced the wearer to deviate slightly from the straight line path adding additional environmental variability to the AG link. The Rx board simulated a radiofrequency wearable device or mobile phone worn by a hiker. The RSSI and SNR values were stored on an SD memory. Multiple flights were completed at three near-ground flying heights: $h=\{3,10,30\}$~m. The wearer walked forth and back many times up to the limits of the forested area. \par
\section{Data Analysis}
\subsection{Radio Range and Packet Delivery Ratio}
The maximum observed radio range and the recorded PDR for the three UAV altitudes are reported in Table~\ref{tab:WoodRangePDR}. A peak in the maximum communication distance is experienced for $h=10$~m, whereas the PDR is~$\gtrsim 90\%$ in all the cases. Beyond the maximum radio range, the PDR suddenly decays to zero, in agreement with the literature on terrestrial LoRa links through forests. The absence of a strong correlation of these parameters with the near-ground altitude of the UAV is coherent with the suburban measurements we reported in~\cite{Bianco22Indirect}. This fact is a further confirmation that an increase in the UAV altitude can both improve or degrade the reception of the packets in near-ground links. Therefore, finding the optimal flying height is not straightforward, unlike the case of scenarios where heightening the drone's altitude can avoid obstacles to establish a LoS ray path and reduce the PL significantly~\cite{Cui19,3GPPTR38901}. To establish a link even with the most distant people without knowing the optimal $h$ in the forest wherein the SaR takes place, the altitude of the aerial drone should vary continuously between the minimum and maximum possible values. This procedure could be exploited, for instance, to broadcast the GPS coordinates of the rescue point.
\begin{figure}[tp]
  \centering
  \includegraphics[width=80mm]{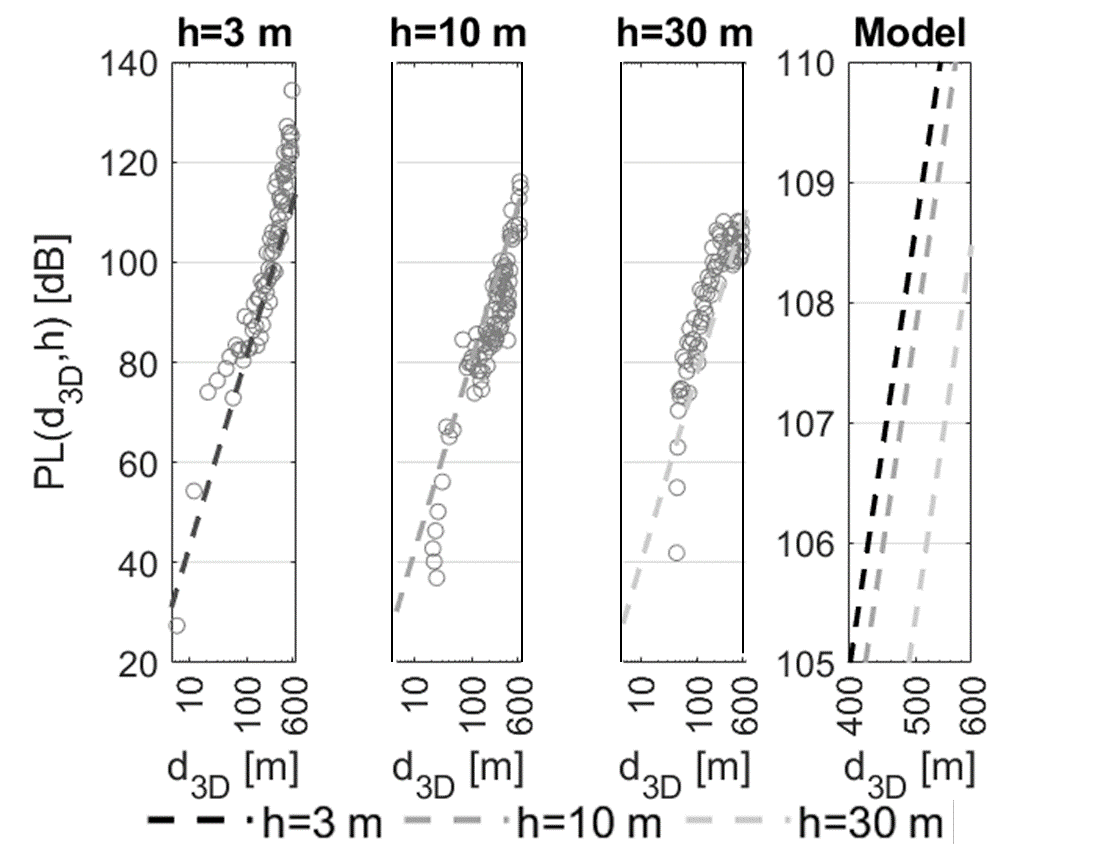}\\
  \caption{Averaged PL measurements and zoomed-in view of the proposed model. The lines derived from the model for different flying heights are coloured in different shades of grey, as in the legend.}
  \label{fig:WoodModel}
\end{figure}
\begin{figure}[tp]
  \centering
  \includegraphics[width=40mm]{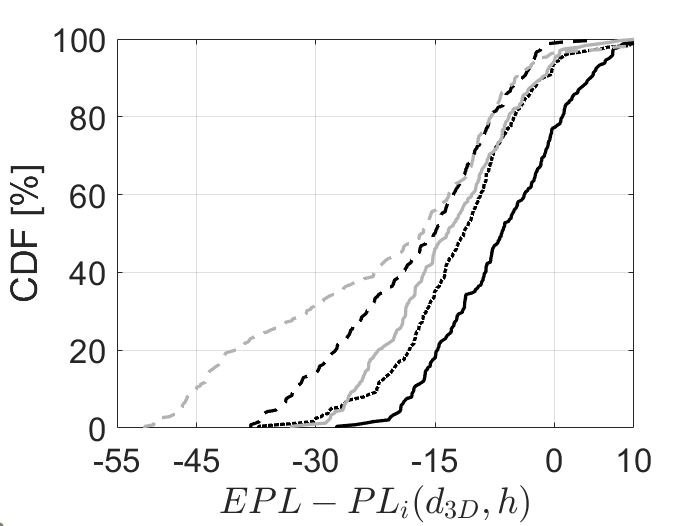}\includegraphics[width=40mm]{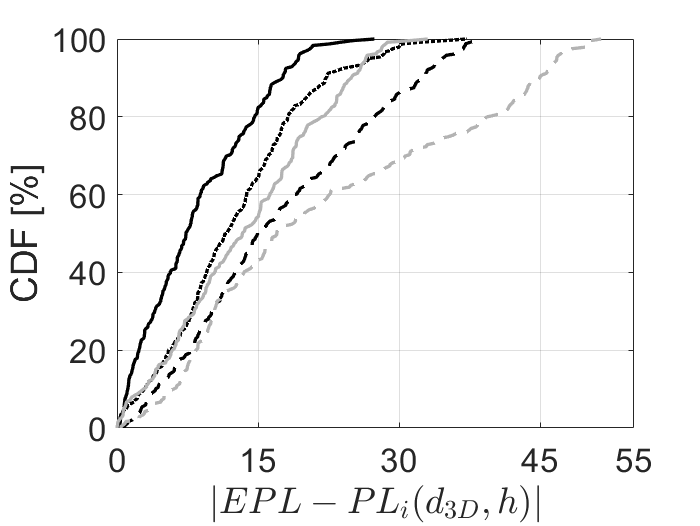}\\
\includegraphics[width=60mm]{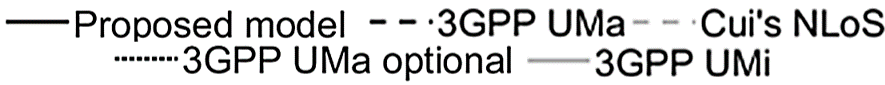}\\
(a) \qquad \qquad \qquad \qquad \qquad (b)
  \caption{CDF of the difference between the EPL by the propagation models and the measured instantaneous PL. (a) Difference and (b) absolute value.}
  \label{fig:ModelComparison}
\end{figure}
\begin{figure}[tp]
  \centering
  \includegraphics[width=40mm]{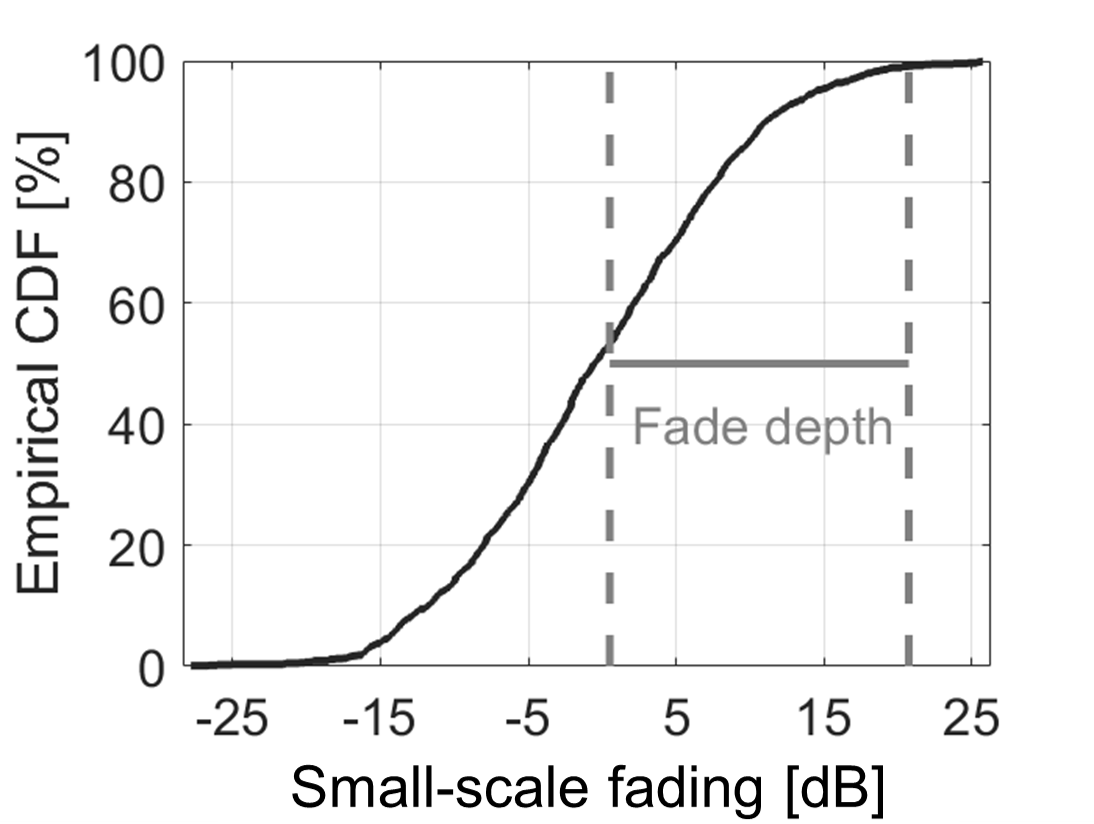}\includegraphics[width=40mm]{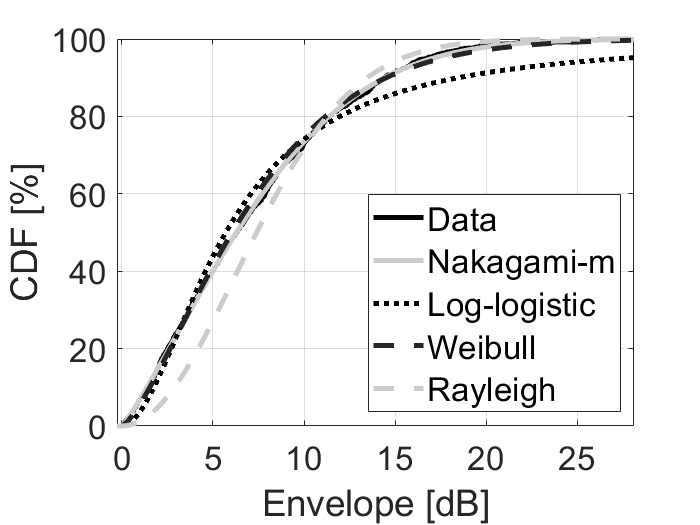}\\
(a) \qquad \qquad \qquad \qquad \qquad  (b)\\
  \caption{(a) Fade depth and (b) statistical distribution of small-scale fading for some standard fits.}
  \label{fig:SmScFading}
\end{figure}
\subsection{Large-scale Fading}
Even if near-ground altitudes were investigated, the mean PL experienced by the signal is decreased when heightening $h$, as it is depicted in Fig.~\ref{fig:WoodModel}. For the sake of benchmarking, the fitting parameters of the EPL model derived in this work are reported in Table~\ref{tab:WoodModel}, together with the parameters of state-of-the-art AG NLoS models whose $\{\gamma,PL\left(1~\textnormal{m}\right),\eta,\sigma_{SF}\}$~values were derived from the literature as detailed at the end of \S\ref{subsec:linkchara}. The resulting altitude impact factor is lower than in urban environments since no direct ray path can ever be achieved by increasing $h$ over the trees. Variable body shadowing contributes to increasing $\sigma_{SF}$ above the benchmark values. The average RSE, calculated for the new model according to (\ref{eq:RSE}), is lower than $4.7\%$, and the statistical validity of the model is thereof proven.\par
%
To evaluate the difference between the EPL by propagation models and the actual $PL_i$ in our scenario, $\{d_{3D},h\}$ parameters were given as input to the models to compute (\ref{eq:PLUAV}) assuming no shadow fading, then the corresponding $PL_{i}(d_{3D},h)$ is subtracted from the EPL for each PL measurement. All the benchmarking models underestimate the fading experienced by the signal (Fig.~\ref{fig:ModelComparison}(a)). The proposed model instead fits the instantaneous PL better than the existing ones, with a reduction in the median error between $4.10$ and $9.66$~dB~(Fig.~\ref{fig:ModelComparison}(b)).
\subsection{Small-scale Fading}
\begin{table}[tp]
  \centering
  \caption{Fade depth and maximum fade.}
\begin{tabular}{l|l||l|l}\label{tab:SmScFadingDepth}
$\mathbf{50}$\bf{\%} \bf{fade}& $0.5$ dB &  $\mathbf{99}$\bf{\%}  \bf{fade}& $20.75$ dB \\
\hline
 \bf{Fade depth} & $20.25$ dB & \bf{Maximum fade} & $21.6$ dB\\
\end{tabular}
\end{table}
The maximum small-scale fluctuation in the signal strength over all the measurements exceeded $30$~dB, as shown by the empirical cumulative distribution function (CDF) of the small-scale fading values in Fig.~\ref{fig:SmScFading}(a). The fade depth is higher than $20$~dB (Table~\ref{tab:SmScFadingDepth}) and hence stronger than NLoS AG models for urban links \cite{Cui19}. The wearer's movements cause swift variations in $G_{Rx}$ and $\chi$, thus deepening the fading.\par
Regarding the statistical distribution, the envelope of the small-scale fading in Fig.~\ref{fig:SmScFading}(b) shows that the small-scale fading is severe, namely, worse-than-Rayleigh~\cite{Matolak11}. In particular, the Rician distribution fit overlaps the Rayleigh fit, thus confirming the continuous absence of dominant LoS components. The best fitting for the fading distribution is determined through maximum log-likelihood and the envelope of the fading \cite{Cui19}. The Nakagami-m distribution best fits the gathered data, with fading figure $\mu=0.64$ and spread parameter  $\Omega=32.27$~\cite{Popovic07}. Hence, multipath scattering is the dominant small-scale propagation phenomenon \cite{Popovic07} due to body movements and dense foliage.
%
%
%
\section{Conclusion}
A body-UAV link through a Mediterranean forest has been evaluated for the first time involving a near-ground UAV reproducing a SaR scenario. The maximum LoRa radio range is about $600$~m, and hard degradation of the signal is observed beyond this limit. The large-scale attenuation is higher than that predicted by state-of-the-art AG NLoS models that are developed for urban environments since they underestimate the median PL by about $4$-$10$~dB. The small-scale fading is about $20$~dB deep, severe, and follows a Nakagami-m distribution. The body-UAV PL is very different from that reported over flat lands (max. radio range~$\sim10$~km)~\cite{Bianco22Numerical} and mountain canyons (max. radio range~$\sim350$~m)~\cite{Bianco22Measurements}, with more relevant small-scale fading, which can stimulate new challenges for localization and emergency communications. Moreover, since the poplars in the Mediterranean forests were young adults, even stronger attenuation is expected if the trees in the forested areas are more ancient or dense~\cite{Guerriero20}. Finally, even though the proposed model was derived from a particular scenario, it can nevertheless be extended to different environmental conditions by introducing proper corrective factors, which can be borrowed from the growing literature on LoRa propagation.\par
\section*{Acknowledgments}
The authors thank professor Leila Guerriero for her helpful comments on the data analysis.
\bibliographystyle{IEEEtran}
\bibliography{LoRa_UAV_Journal_14.bib}

\end{document}